\documentclass{IEEEtran}

\pdfoutput = 1

\usepackage{algpseudocode}
\usepackage{algcompatible}
\usepackage{algorithm}
\usepackage{subfigure}
\usepackage{graphicx}
\graphicspath{{ToConvert/}}

\newtheorem{theorem}{Theorem} 

\renewcommand{\thepage}{} 
\begin{document}
%
\title{TRASA: TRaffic Aware Slot Assignment Algorithm in Wireless Sensor Networks}


{
\author{\IEEEauthorblockN{Ichrak Amdouni and Pascale Minet}\\
\IEEEauthorblockA{INRIA Rocquencourt\\
78153 Le Chesnay cedex, France\\
Email: ichrak.amdouni@inria.fr, pascale.minet@inria.fr}}
}

\maketitle

\begin{abstract}
In data gathering applications which is a typical application paradigm in wireless sensor networks, sensor nodes may have different traffic demands. Assigning equal channel access to each node may lead to congestion, inefficient use of the bandwidth and decrease of the application performance. In this paper, we prove that the time slot assignment problem is NP-complete when $p-hop$ nodes are not assigned the same slot, with $1 \le p \le h$ for any strictly positive integer $h$. We propose {\em TRASA}, a TRaffic Aware time Slot Assignment algorithm able to allocate slots to sensors proportionally to their demand. 
We evaluate the performance of {\em TRASA} for different heuristics and prove that it provides an optimized spatial reuse and a minimized cycle length.
\end{abstract}


%
\IEEEpeerreviewmaketitle

\section{Context and motivations}

A typical application in Wireless Sensor Networks (WSNs) is data gathering. Sensor nodes are deployed in the region of interest to periodically collect and report sensed data to a sink node.
To achieve this many-to-one communication, sensor nodes form a data gathering tree rooted at the sink.
This communication paradigm is also known as "convergecast"~\cite{FASTDATA_Collec}.

Depending on the application requirement, different objectives have to be taken into account in the design of MAC protocols.
For instance, reliability is required mainly in critical applications like fire detection. In this disaster scenario, reducing the end-to-end delays is also a major requirement. Energy-efficiency is an important concern specially when sensors are operating for a long period of time.
It has been proved that, under heavy traffic conditions, contention-free medium access based on TDMA outperforms contention-based protocols.

The main task of TDMA-based MAC protocols is to assign time slots to sensor nodes and to schedule the medium access based on these slots. 
The transmission schedule allows nodes to send and receive data packets without collision. 
Further, any node can enter into sleep mode during inactive periods, thus achieving low duty cycle and saving energy.
Although the use of TDMA requires synchronization between cooperating sensors, it is an efficient way
of mitigating the limitations of CSMA based networks.
However, the performance of the TDMA-based medium access protocols may dramatically decrease if the TDMA parameters are not aware of the application requirements~\cite{TDMA-DESIGN}; for example, a very long frame may increase the latency. Moreover, traditional MAC protocols tend to give nodes equal channel access, while the sensor traffic demands may differ.
Indeed, nodes close to the sink forward more data than leaf nodes in the data gathering tree. This is the "funneling effect"~\cite{FUNN}.
Allocating equal numbers of time slots to sensor nodes may lead to congestion, packet loss, and inefficient use of the bandwidth. Consequently, channel access should be proportional to the sensor demand.  
That is why we investigate in this paper the {\em Time Slot Assignment} problem, denoted \textit{TSA}, regarding the application requirements. We propose \textit{TRASA}, \textit{TRaffic Aware Slot Assignment algorithm} for WSNs. Assuming a sensor network where each node has a specific number of packets to transmit to its parent in the data gathering tree, {\em TRASA} assigns each node a number of slots proportional to its traffic demand, and schedules its activity. Moreover, {\em TRASA} allows the allocation of slots to nodes with heterogeneous demands. Consequently, the algorithm addresses the funneling effect and ensures a fair medium access. 
{\em TRASA} builds a schedule and ensures that data reach the sink in one cycle. {\em TRASA} takes into account both tree communication links and other interfering links. Via simulations, we evaluate the impact of interfering links on {\em TRASA} algorithm, and show that as the spatial reuse decreases, the schedule length increases. 
Since the time slot assignment problem is NP-complete as we will prove hereafter, {\em TRASA} relies on an heuristic determining the priority of each node. We compare this heuristic with another heuristic and justify that {\em TRASA} ensures an optimized schedule length and a good tradeoff between the schedule length, the average end-to-end delay and the maximum buffer size required by each node.\\  
The remainder of this paper is organized as follows.
In section~\ref{stateArt:sec}, we provide a state of the art about the {\textit TSA} problem. In section~\ref{pbDef:sec} we describe the assumptions adopted and define the problem statement. In section~\ref{complexity:sec}, we prove that the \textit{TSA} problem is NP-complete, assuming that two nodes that are $h$-hop away, with $h$ a positive integer $>1$ interfere.
Section~\ref{TRSADesc:sec} describes the \textit{TRASA} algorithm and its properties. In section~\ref{performance:sec}, we present the performance evaluation of \textit{TRASA} and conclude in section~\ref{conclusion:sec}.


\section{State of the art}\label{stateArt:sec}
\begin{figure*}[!ht]
\begin{center}
	\includegraphics[width=5.7in]{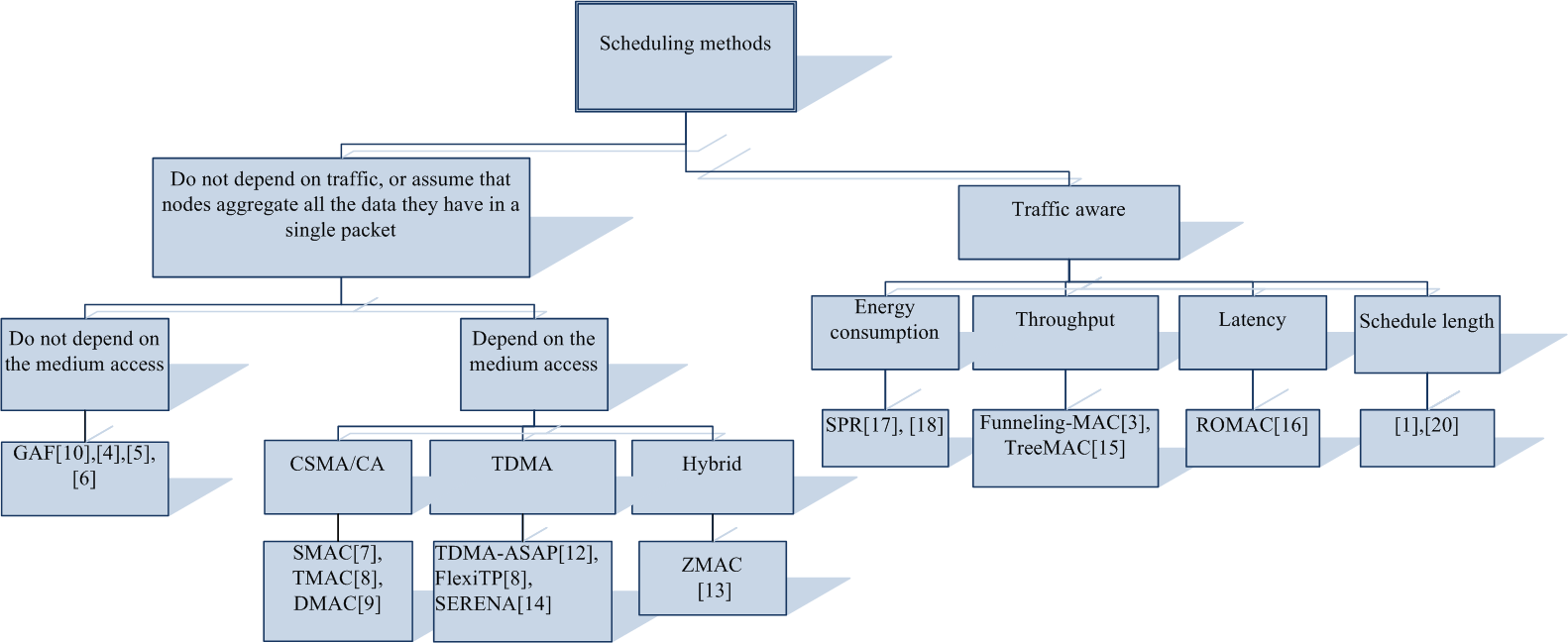} 
	\caption{Taxonomy of time slot assignment techniques.\label{taxonomy}} 
\end{center}
\end{figure*}

Despite the existence of a variety of scheduling schemes, few of them allocate a number of slots proportional to node demand. In this section, we present a state of the art of existing schemes classified according to their awareness of the traffic demand. This classification is illustrated in Figure~\ref{taxonomy}. 

\noindent $\circ${\textbf{ Protocols that do not depend on the traffic demand, or assume that nodes aggregate all the data they have to transmit in a single time slot}:}\\
With these protocols, any node receives exactly one slot.  For space reasons, we do not detail here this class of algorithms.\\
\noindent$\circ${\textbf{ Traffic aware protocols:}\\}
In these protocols, the number of slots received by a node depends on its traffic demand. We present some examples of {\textit TSA} algorithms classified according to their main objective. Notice that one algorithm may satisfy more than one objective.\\
\textbf{1) Maximizing throughput of received data at the sink}:\\
 In~\cite{FUNN}, Gahng et al. presented Funneling-MAC to mitigate the funneling effect and boost application fidelity in WSNs. Funneling-MAC is hybrid, TDMA is used within small number of hops from the sink (called the intensity region), and CSMA/CA elsewhere. The sink is responsible for the scheduling of the nodes in this intensity region. 
Funneling-MAC allows the slot reuse between nodes distant more than 2 hops.
The main goal of TreeMAC~\cite{Tree-MAC} is to achieve high throughput in real-time high-data-rate WSNs. Authors proved they achieve at least one-third of the optimum throughput assuming reliable links. In TreeMAC, slots are assigned from the root to the leaves, making it not suitable for large-scale networks. TreeMAC allows a slot reuse between nodes belonging to the same tree branch, but not between the subtrees. From energy point of view, authors show that TreeMAC outperforms CSMA and Funneling-MAC~\cite{FUNN}.
However, nodes in Tree-MAC are assigned disjoint transmitting time slots, which means that if the node can sleep between its transmitting slots, it should awake up many times to transmit and receive its data. The transitions between radio states are energy-consuming, so should be reduced.\\
\textbf{2) Minimizing latency:}\\
ROMAC~\cite{ROMAC} is a localized QoS-aware for high-fidelity data center sensing networks. Like TreeMAC, ROMAC divides the TDMA cycle into frames, each frame being composed of three slots and allocates slots to nodes proportionally to their demand. Compared to TreeMAC, ROMAC enhances the localization aspect, since each node can locally determine its frame and its time slots without relying on its parent like in TreeMAC. The node with identifier $i$ transmits its own data in the frame $i$, and transmits the packets of any node $j$ in its subtree in the
$j^{th}$ frame. Unlike TreeMAC and Funneling-MAC, ROMAC automatically adapts to routes changes, any node updates its set of frames every TDMA cycle. 
Consequently, ROMAC achieves lower delivery latency than these two protocols as the network size increases.\\
\textbf{3) Minimizing schedule length:}\\
Incel et al.~\cite{FASTDATA_Collec} aimed at reducing the delays of data collection by minimizing the schedule length. 
They studied scheduling nodes where each node generates a packet at the beginning of the TDMA frame, in convergecast scenario keeping tree interference links only. They proved that the lower bound on the schedule length is given by: (1) the maximum node degree when packet aggregation at each intermediate node is considered, and (2)  $max(2n_k-1,N)$ where $n_k$ is the maximum number of descendants of the sink children, otherwise.
For this second case, authors proposed \textit{Local-TimeSlotAssigment} algorithm in which the sink schedules an edge having the highest remaining number of packets, and any parent node with an empty buffer selects one child whose buffer is not empty at random respecting tree interfering links. This way, they ensure parallel transmissions among multiple branches of the tree, and keep the sink busy in receiving packets. In~\cite{Ergen10}, authors proposed algorithms based on coloring.
Two centralized solutions are described. First, in the node-based scheduling any slot is shared between nodes with the same color and any other node that does not conflict with them. Second, in the level-based scheduling, conflicts are defined between levels: the same color is assigned to two levels if they do not contain any couple of conflicting nodes. For each color, non conflicting nodes from all the levels having this color share the same slot. Besides other nodes that do not conflict with already scheduled nodes are scheduled simultaneously. This algorithm suffers from the energy waste because of the radio switches between active/sleep states.\\
\textbf{4) Minimizing the energy consumption:}\\ 
In~\cite{SPR}, Turau et al. proposed SPR, to schedule each path in the routing tree separately. The slots assigned to any node are the union of its slots in each path. Hence, the spatial reuse of time slots is restricted to a common path, which makes SPR not efficient in terms of schedule length for networks where the average number of children is high. Another example in this category is given in~\cite{scheSchemes07} where the energy efficiency is achieved by reducing the number of switches between the active and sleep states.

\section{Time slot assignment problem}\label{pbDef:sec}
In this section we present the assumptions we adopt to study the time slot assignment problem {\em TSA}.

\subsection{Assumptions and system model}\label{assump:sec}
\noindent \textbf{$\circ$ Network model:} The network is modeled as a graph $G=(V,E)$, $V$ is the set of vertices, and $E$ is the set of edges representing the communication links. $T$ is a spanning tree of $G$ rooted at the sink node. We adopt the unit disk model, where nodes are modeled as a set of points in the 2-dimensional plane. Besides, there is no message losses. A node $u$ receives a message sent by another node $v$, if and only if, their distance is lower than a given uniform transmission range $R$. 
Nodes $u$ and $v$ are then 1-hop neighbors. For any integer $h>1$, any two nodes $u$ and $v$ are $h$-hop neighbors if and only if $u$ is $(h-1)$-hop away from a 1-hop node of $v$.\\
\textbf{$\circ$ Interfering links:} Usually, in a data gathering application, the communications of a node are limited to its parent and its children. However, any node may have communication links with other nodes. If these links are not taken into account in node scheduling, collisions may occur. Hence, for a given node, we include all its interfering links, those in the tree and other ones. 
In the literature, solutions to the {\em TSA} problem for data gathering applications tend to limit the interferences to 2 hops. However, in real wireless networks, interferences may exceed this distance. 
Consequently, to generalize our study, we assume that two nodes that are $p$-hop away with $1 \le p \le h$ are interfering. $h$ being a given positive integer $>1$ which is a parameter of our algorithm. Hence, we denote the studied problem $h\_TSA$.\\
\textbf{$\circ$ Application data:} We consider a data gathering application. In each TDMA cycle, each node except the root has its own data to transmit in addition to the data received from its children. Some nodes (for example, the children of the sink), need more than one slot to transmit their data.

 
\subsection{Problem statement}
We study the slot assignment problem under the assumptions introduced in Section~\ref{assump:sec}.
The slot assignment problem consists in assigning slots to nodes in $G$, such that no two nodes that are $p$-hop away with $1 \le p \le h$ are scheduled in the same slot while minimizing the schedule length. Besides, this scheduling must ensure that each node transmits its own packets, and the packets generated in its subtree.

\subsection{Lower and upper bounds}
\begin{theorem}
The number of slots required by the $TSA$ problem meets:\\
\begin{equation}
\mathrm{Number~of~nodes}-1 \le \mathrm{Number~of~slots} \le \sum_{\mathrm{node~u}}\mathrm{depth(u)},
\end{equation}
where depth(u) is the depth of node $u$ in the tree.
\end{theorem}
\begin{IEEEproof}
In the best case, all the sink children transmit their packets in sequence. Indeed, there is no possible spatial reuse between sink children. Hence the lower bound.
In the worst case, each slot corresponds to a single packet transmission. The number of slots needed by any packet generated by any node $u$ at depth $d$ in the data gathering tree is equal to $d$. Hence the upper bound.
\end{IEEEproof}
\section{Complexity of the $h\_TSA$ problem}\label{complexity:sec}
It has been proved in \cite{Choi07} and \cite{Ergen10} that the {\em TSA} problem is NP-complete. 
In this paper, we generalize the study and prove that the $h\_TSA$ problem is NP-complete for any  positive integer $h$ where any two nodes that are less than or equal to $h$-hop away are not scheduled simultaneously.
\begin{theorem}
The $h\_TSA$ problem, for any positive integer $h\ge1$ is NP-complete.
\end{theorem}
\begin{IEEEproof}
The decision problem of {\em h\_TSA} is given by: {\em Can nodes in a graph $G$ be assigned $S$ time slots ($S$ is a positive integer) during which they can transmit their data to a sink node, such that any two nodes that are $p$-hop away, with $1 \le p \le h$, are not assigned the same time slot?}\\
To prove that the decision problem of {\em h\_TSA} is NP-complete, we 
use the knowledge that the {\em h\_Color} problem is NP-complete~\cite{WMNC2011}.
In~\cite{WMNC2011}, the {\em h\_Color} problem is defined as {\em coloring a graph with the smallest number of colors (a color is represented by an integer) such that any two nodes that are $p$-hop away with $ 1\le p \le h$, do not have the same color. In addition, the color assigned to any node is smaller than the color assigned to its parent in the data gathering tree.}\\
We need to prove that finding a solution to {\em h\_TSA} is equivalent to finding a solution to {\em h\_Color}.\\

Let $G$ be a connected, undirected graph, and $T$ its spanning tree.
Notice that the construction of $T$ can be done in polynomial time. 
 Each node $u$ in $G$ has traffic demand $d_u$.  
Let $C_G$ be a coloring of $G$ solving the \textit{h\_Color} problem and using the colors $c_1,c_2,\ldots, c_{max}$. Let $N_i$ be the set of nodes having the color $c_i$. We can build a slot assignment for nodes in $G$ as follows. We sort the colors by increasing order. For the smallest color $c_i$ not yet considered, we add to the cycle a number of slots equal to the maximum traffic demand of nodes in $N_{i}$, denoted $max_d{_i}$. Then, nodes in $N_{i}$ are scheduled during these slots, each one has a number of slots equal to its traffic demand. Consequently:
\begin{enumerate}
\item This scheduling is conflict-free, nodes that are scheduled simultaneously have the same color, and hence do not interfere.
\item The scheduling allows each node to transmit all the data it has since it is assigned a contiguous number of slots equal to its demand. Besides, this scheduling ensures that each parent node accesses the medium after all its children, because it has a higher color than them, and slots are assigned to nodes having the smallest colors first.
\item The number of slots used is equal to $S = \sum_{i=1}^{c_{max}} max_d{_i}$.
\end{enumerate}

Figure~\ref{ComplexityFig} illustrates the proof.
\begin{figure}[h]
	\subfigure[]{\includegraphics[width=1.3in]{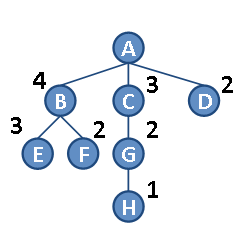}\label{graph}}
    \subfigure[]{\includegraphics[width=2in]{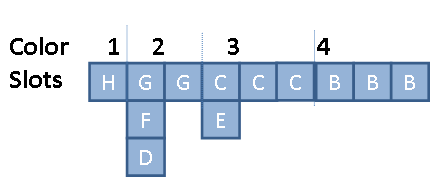}\label{slots}}
\caption{An example illustrating the proof of the NP-completeness of {\textit TSA} problem.\label{ComplexityFig}}
\end{figure}

Inversely, given a scheduling of $G$, composed of $S$ slots, we need to color the graph according to the {\em h\_Color} problem. 
We denote $S_s=s_1,s_2,\ldots,s_s$ the sequence of slots such that during each slot $s_i$ at least one node is scheduled for the last time in the TDMA cycle. Then, for each slot $s_i$ in $S_s$ we assign the color $i$ to all uncolored nodes occupying $s_i$ such that $s_i$ is their last time slot. 
Consequently, we get a number of colors bounded by the cardinal of $S_s$.
The same color is assigned to nodes scheduled in the same slot, so the coloring is collision-free. Further, since the last slot of each parent can not be scheduled before the last slot of any of its children, each parent has a color greater than the color of its children.\\
Hence the theorem.
\end{IEEEproof}

\section{TRASA Algorithm: design and properties}\label{TRSADesc:sec}
\subsection{Overview and algorithm}
\begin{figure}[h]
\begin{center}
\includegraphics[width=1.8in]{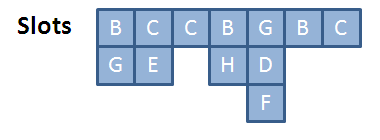}
\caption{An example of slot assignment by TRASA.\label{TRASAExample}}
\end{center}
\end{figure}

The requirements of data gathering applications are various: a short TDMA cycle, a high throughput at the sink, small delays and low energy consumption. {\em TRASA} addresses these issues by maximizing the slot reuse in a cycle. Moreover, by favoring nodes with a high number of descendants, {\em TRASA} tends to minimize the TDMA cycle length and then saves energy.
Figure~\ref{TRASAExample} shows the slots computed by {\em TRASA} applied to the graph in Figure~\ref{graph}.

As the $h\_TSA$ problem is NP-complete, heuristics must be used. In {\em TRASA}, sensor nodes are sorted according to their priority. The priority of a node is given by the number of its descendants in the data gathering tree. Which means that the nodes close to the sink have the highest priority. 
\begin{algorithm}
\caption{TRASA algorithm.}\label{algo}

\begin{algorithmic}[1]
\STATE Input: a connected graph $G$, and its spanning tree $T$, where each node $u$ has $d_u$ packets to transmit
\STATE Output: $Cycle$: the TDMA frame with the assigned slots per node
\STATE $Cycle.endingSlot = 0$ /*the last slot assigned during the current iteration*/
\STATE $Cycle.lastEndingSlot = 0$ /*the first slot assigned during the current iteration*/
\WHILE {(There is at least one node having data to transmit)}
\STATE $N_s$ = List of nodes having data to transmit sorted according to their priority
\STATE $u$ = node with the highest priority in $N_s$
\STATE $Cycle$.addSlot($d_u$) /*$d_u$ slots are added to $Cycle$*/
\STATE $Cycle.lastEndingSlot = Cycle.endingSlot$
\STATE $Cycle.endingSlot$ += $d_u$
\STATE $u$.schedule.add(($Cycle.lastEndingSlot$,$d_u$)) 
\STATE $u$.demand -= $d_u$
\STATE $u$.parent.demand += $d_u$
	\FOR {node $v$ in $N_s$}
	\IF {($v \ne u$) \&\& (v has $d_v\ne0$ packets) \&\& (v does not interfere with nodes occupying the slots between $lastEndingSlot$ and $endingSlot$)} 
		   \STATE \textit{d=Cycle.endingSlot-Cycle.lastEndingSlot}
		   \IF {($d_v > d$)} 
		   \STATE $Cycle$.add($d_v-d$)
		   \STATE $Cycle.endingSlot$+=$d_v-d$
		   \ENDIF
		   \STATE $v$.demand -= $d_v$
		   \STATE $v$.parent.demand += $d_v$
	   
	\ENDIF 
	\STATE $v$.schedule.add(($Cycle.lastEndingSlot$, $d_v$))
	\ENDFOR 
\ENDWHILE
\end{algorithmic}
\end{algorithm}
 The reasons for this choice are threefold: (1) Nodes with many descendants need to forward more packets than others, so scheduling them as soon as they have data avoids the congestion, and allows them to use less buffers. (2) The children of the sink are interfering, only one child can be active in a slot. They are the most determining factors of the schedule length (their number represents the lower bound of the number of slots). We believe that scheduling them first achieves a higher flexibility. (3) The probability of the sink to be in receive state each time slot is high, and similarly the throughput measured at the sink increases.\\
 The pseudo-code of {\em TRASA} is given by algorithm~\ref{algo}.\\
 Once the nodes are sorted, the node $u$ with the highest priority is given a number of slots equal to its packet demand. This is because we assume for simplicity reasons that a time slot contains only one packet. Notice that {\em TRASA} is able to take into account sensor nodes with heteregeneous data rates. 
\textit{TRASA} assigns node $u$ $d_u$ slots between indices $lastEndingSlot$ and $endingSlot$.
Aiming at reducing the schedule length, {\em TRASA} achieves an optimized slot reuse. That is, a new time slot is added to the TDMA frame, if and only if, nodes having remaining data, could not be scheduled in the last time slot. The only limiting factor of the parallelism is the interferences which have to be considered to achieve a collision-free schedule.
So to schedule other nodes simultaneously with the node $u$, {\em TRASA} iterates on the nodes having data to transmit according to the descending order of their priority. If a node $v$ has packet(s) to send and is not interfering with nodes already scheduled between $lastEndingSlot$ and $endingSlot$, $v$ is scheduled in parallel with these nodes. The number of slots assigned to $v$ is equal to its demand $d_v$. Notice that $d_v$ may be higher than $endingSlot-lastEndingSlot$, in this case, the TDMA cycle is extended. \textit{TRASA} stops when all nodes have transmitted all data they generated, and all data they received from their children.

\subsection{TRASA properties}
In this section, we discuss the properties of {\em TRASA}, and explain how it takes into account various performance criteria.\\
\textbf{ P1. Fair Access:} The first key design of {\em TRASA} is to ensure a fair medium access to all nodes. A time slot being associated to a unique packet, a node has as many slots as packets to transmit. In a data gathering tree, any node traffic demand is the sum of the packets it generates and the packets generated in its subtree. {\em TRASA} assigns each node as many slots as its traffic demand. Hence, {\em TRASA} addresses the funneling effect and avoids congestions.\\
\textbf{ P2. Optimized spatial reuse:} {\em TRASA} assigns slots to nodes in the order of their priority. When any node is scheduled, all non interfering nodes are scheduled simultaneously. Hence, the slot reuse is optimized, it is not restricted to a common branch, like~\cite{Tree-MAC}, or inter-branches, like~\cite{SPR}. This spatial reuse provides an efficient use of the bandwidth.\\
\textbf{ P3. Optimized schedule length:} In {\em TRASA}, the TDMA frame is extended by a time slot if and only if the nodes with a packet demand cannot be scheduled in the last added slot, because of the interferences. This behavior reduces considerably the schedule length. For instance, applied to the graph of Figure~\ref{graph} for $h=2$, \textit{TRASA} gives a cycle length of 7 slots as illustrated by Figure~\ref{TRASAExample}. The cycle length would be 12 slots with TreeMAC. \\
\textbf{ P4. Optimized energy consumption:} Minimizing the number of slots in a cycle reduces the activity periods and allows nodes to save energy. Besides, when any node is scheduled, it is allowed to send all data it has, which reduces the switches between the active and sleep states which are energy-consuming. With {\em TRASA}, any node is awake during its slots to transmit data and the slots of its children to receive data they transmit, it sleeps the remaining time to save energy.\\
\textbf{ P5. Optimized buffer usage:} In WSNs, nodes have low capacity of storage. Accumulating many packets in a node may lead to buffer overflow and packet losses. In {\em TRASA}, the buffer usage is balanced during the TDMA frame. 
Indeed, only nodes having packets to transmit compete for a given time slot. Since the priority used is the number of descendants, the priority of any node is higher than the priority of all its children. Consequently a parent having data to transmit is in general scheduled before its children, so does not accumulate much data.\\
\textbf{ P6. Optimized delays and throughput:} The optimized slot reuse of {\em TRASA} can be seen as a parallel progress of packets toward the sink, and in a balanced way between tree levels and subtrees, as much as allowed by the interference constraints. Consequently, the end-to-end delays are reduced. Furthermore, nodes close to the sink have the highest priority. This increases the probability that  the sink receives data in a time slot, and hence enhances the data throughput. By construction of the cycle, {\em TRASA} guarantees that all data generated by nodes reach the sink in one cycle.

\section{Performance evaluation}\label{performance:sec}
\subsection{Simulation setup}
Using a graph generator, we generate random graphs for a given number of nodes deployed in a 2D area ($1m$x$1m$), the transmission range is set to $0.4m$. The number of nodes ranges from $20$ to $100$. For each graph, we build a spanning tree defined by a maximum number of children per node. Notice that if the graph is disconnected, the construction of the tree is not possible.
The corresponding graph is then not considered in the simulation process.
The tree is built by adding children nodes consecutively. The first added node is the root. Any neighbor $u$ of the root selects it as a parent if the maximum number of children (which is set to $3$) is not reached by the sink. Otherwise, such a node $u$ tries to associate to another node. This process continues until all nodes have a parent. Notice that implicitly, each node selects the parent with the minimum number of hops toward the sink, provided that this parent can accept further children. 
We assume that interferences are limited to $2$-hops ($h=2$), as usually assumed in {\textit TSA} problems.
Each node generates only one packet, and forwards packets it receives from its children. For simplicity reasons, we also assume that a slot contains only one data packet. 
Each result is the average of 40 simulation runs.
\subsection{Simulation results}
 
\subsubsection{Choice of the heuristic}
\begin{figure}[!h]
	\begin{center}
	\subfigure[]{\includegraphics[width=1.662in, angle=270]{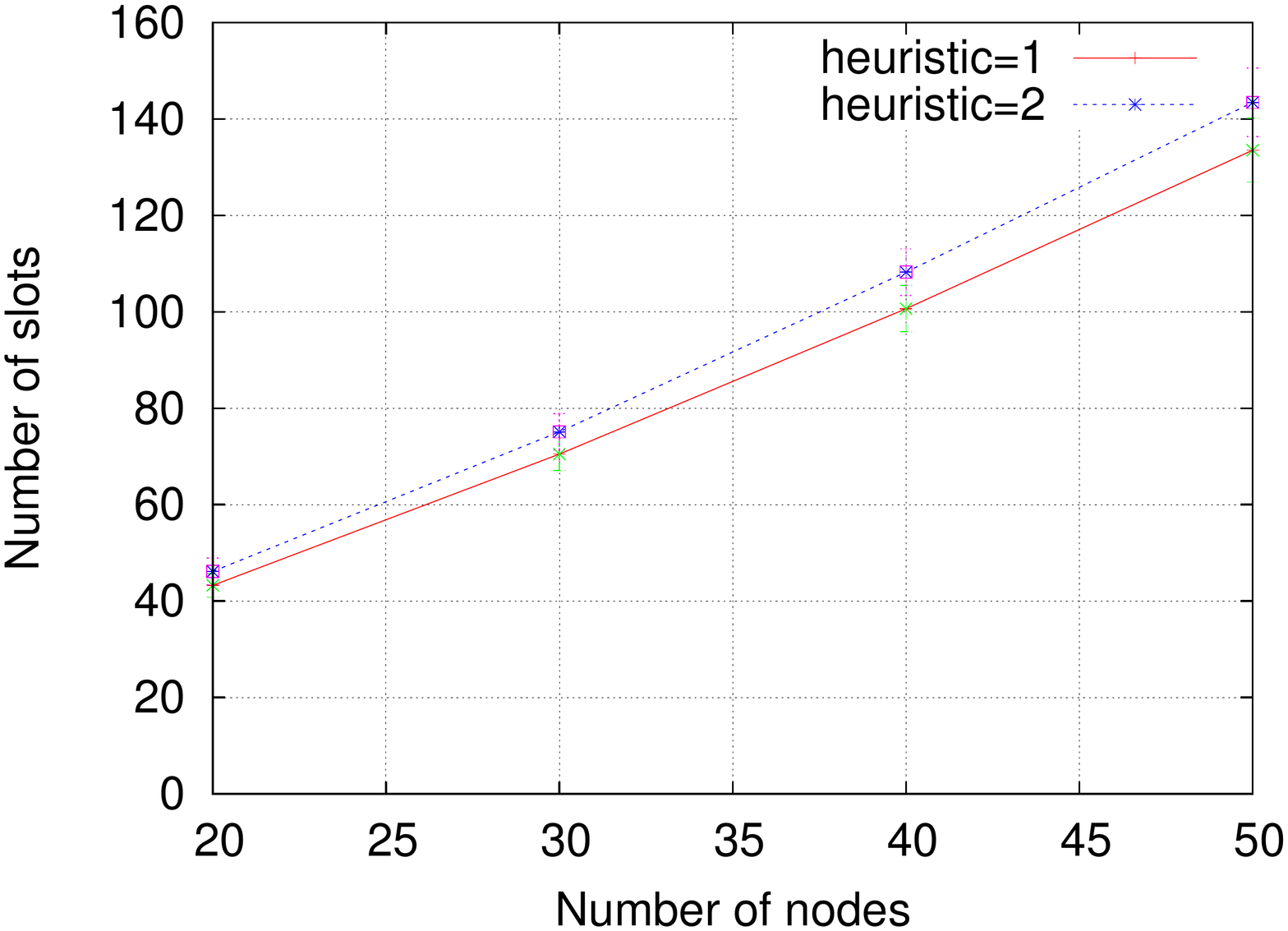}\label{nbSlotH}}
	\subfigure[]{\includegraphics[width=1.982in, angle=270]{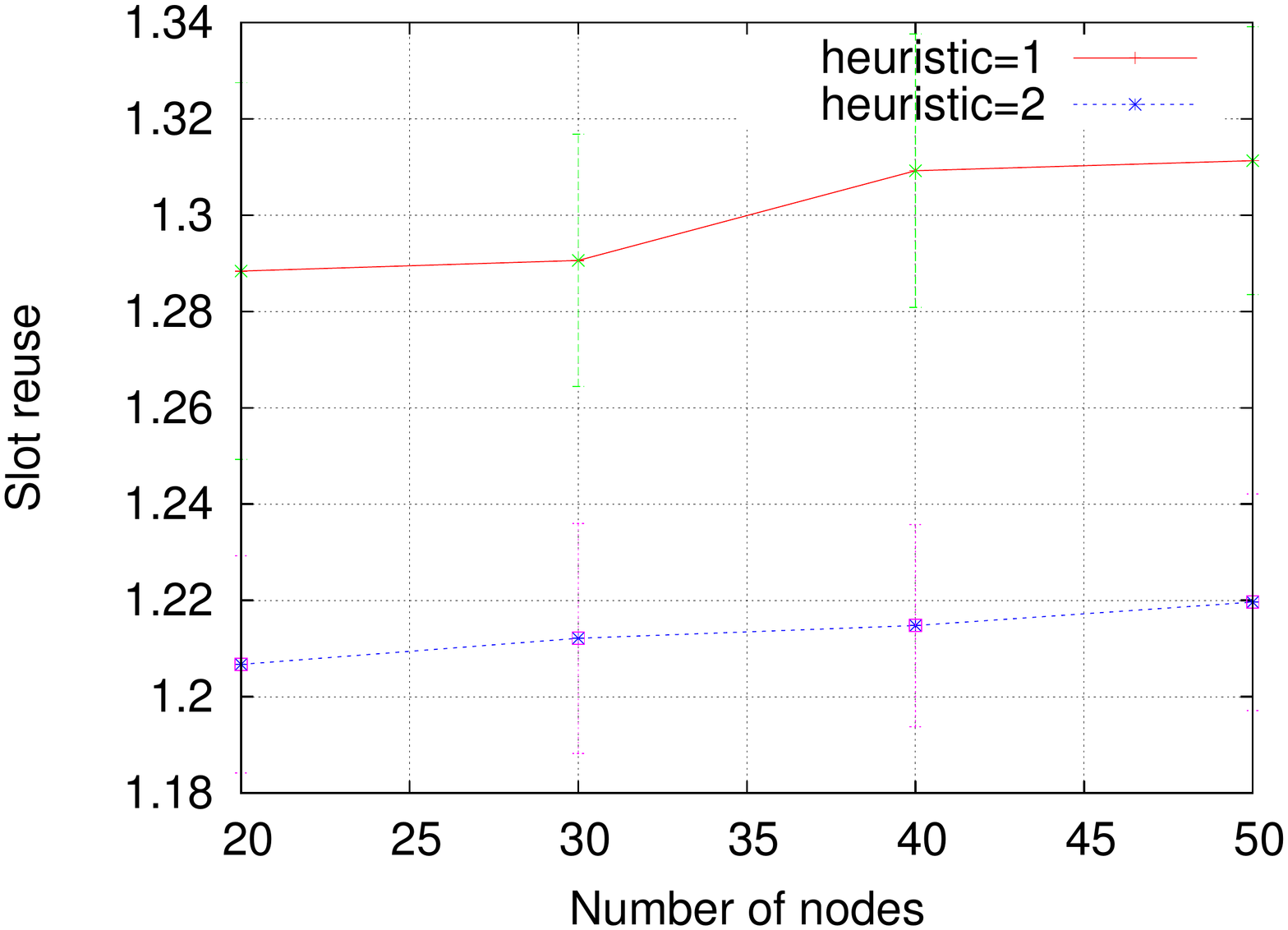}\label{slotReuseH}}
	\subfigure[]{\includegraphics[width=1.662in, angle=270]{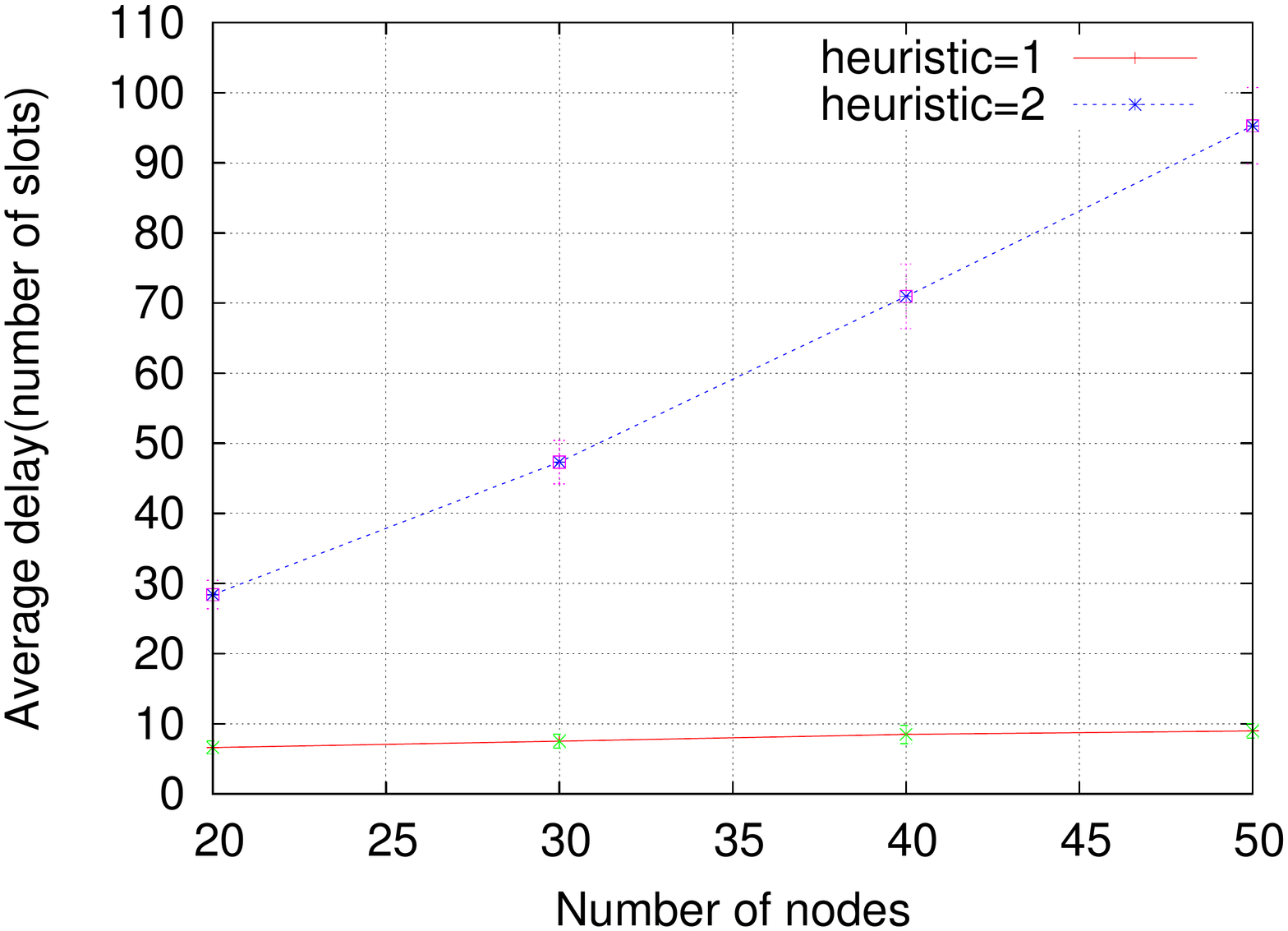}\label{avgDelayH}}
	\subfigure[]{\includegraphics[width=1.982in, angle=270]{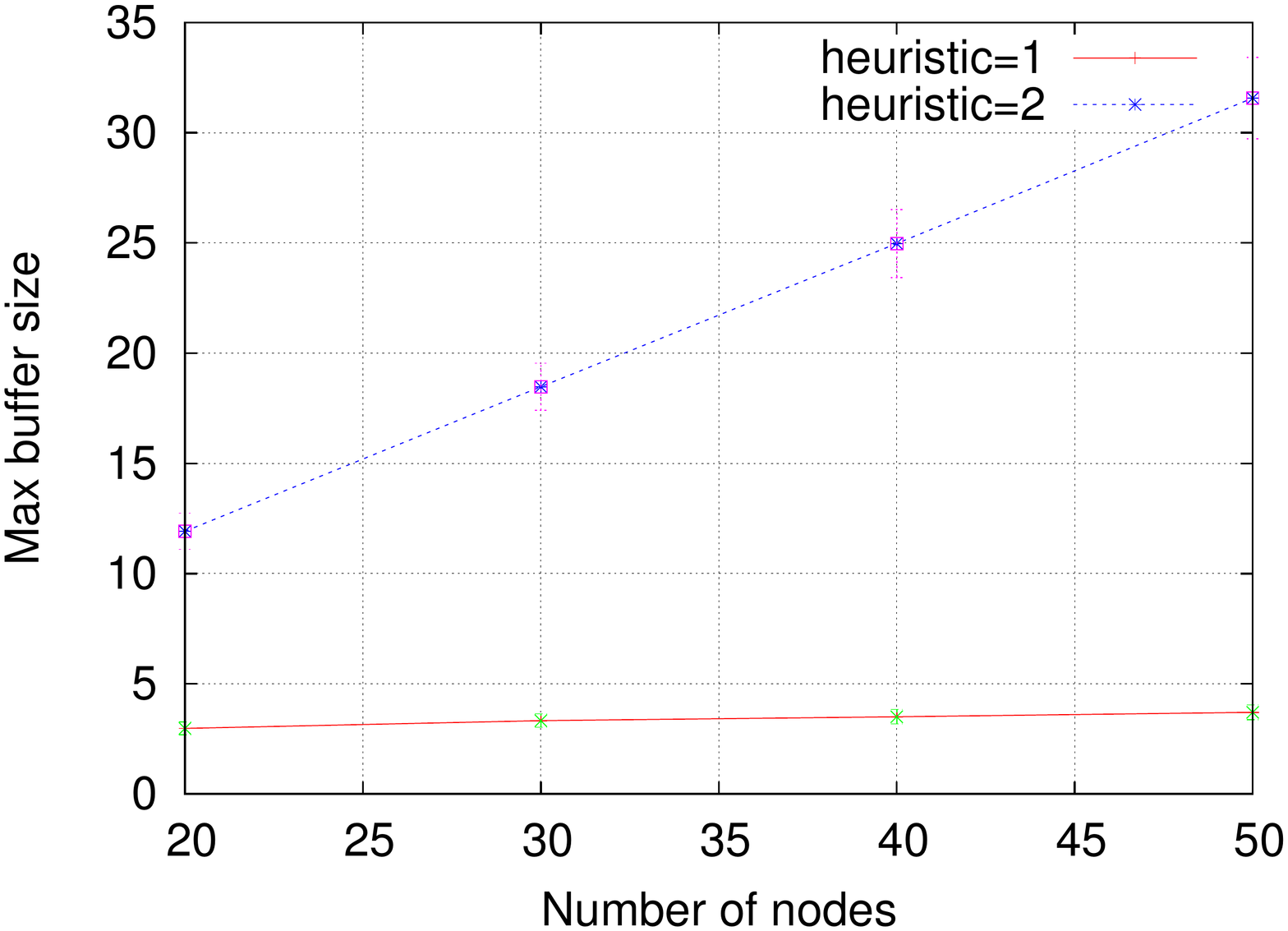}\label{maxBufferSizeH}}
	\caption{Impact of the heuristic choice on: (a) Slot number; (b) Slot reuse; (c) Average delay, (d) Maximum buffer size.} 
	\end{center}
\end{figure}

In {\em TRASA}, nodes are sorted according to their priority which is given by their number of descendants. This heuristic is denoted $ heuristic=1$. 
To validate this design choice, we compare this heuristic with another one, denoted $heuristic=2$, where nodes with the highest number of descendants have the smallest priority.  
Figure~\ref{nbSlotH} shows that $heuristic=1$ outperforms $heuristic=2$ in terms of slot number. 
This is because the slot reuse is increased, as illustrated in Figure~\ref{slotReuseH}.
Figure~\ref{maxBufferSizeH} shows that the maximum size of buffers required by a node in $heuristic=1$ is lower than in $heuristic=2$. Indeed, when leaf nodes are scheduled first, a node is likely to accumulate packets before being able to transmit them. Whereas $heuristic=1$ gives opportunity to nodes to send data as soon as possible, which considerably reduces average end-to-end delays as illustrated in Figure~\ref{avgDelayH}. However, nodes may access the medium more frequently during one TDMA cycle, which adds an energy cost when the number of switches between active and sleep state increases.
For these reasons, $heuristic=1$ is preferable and is selected in the following.\\
\subsubsection{Impact of interference links}
\begin{figure}[!ht]
\centering
\subfigure[]{\includegraphics[width=1.662in, angle=270]{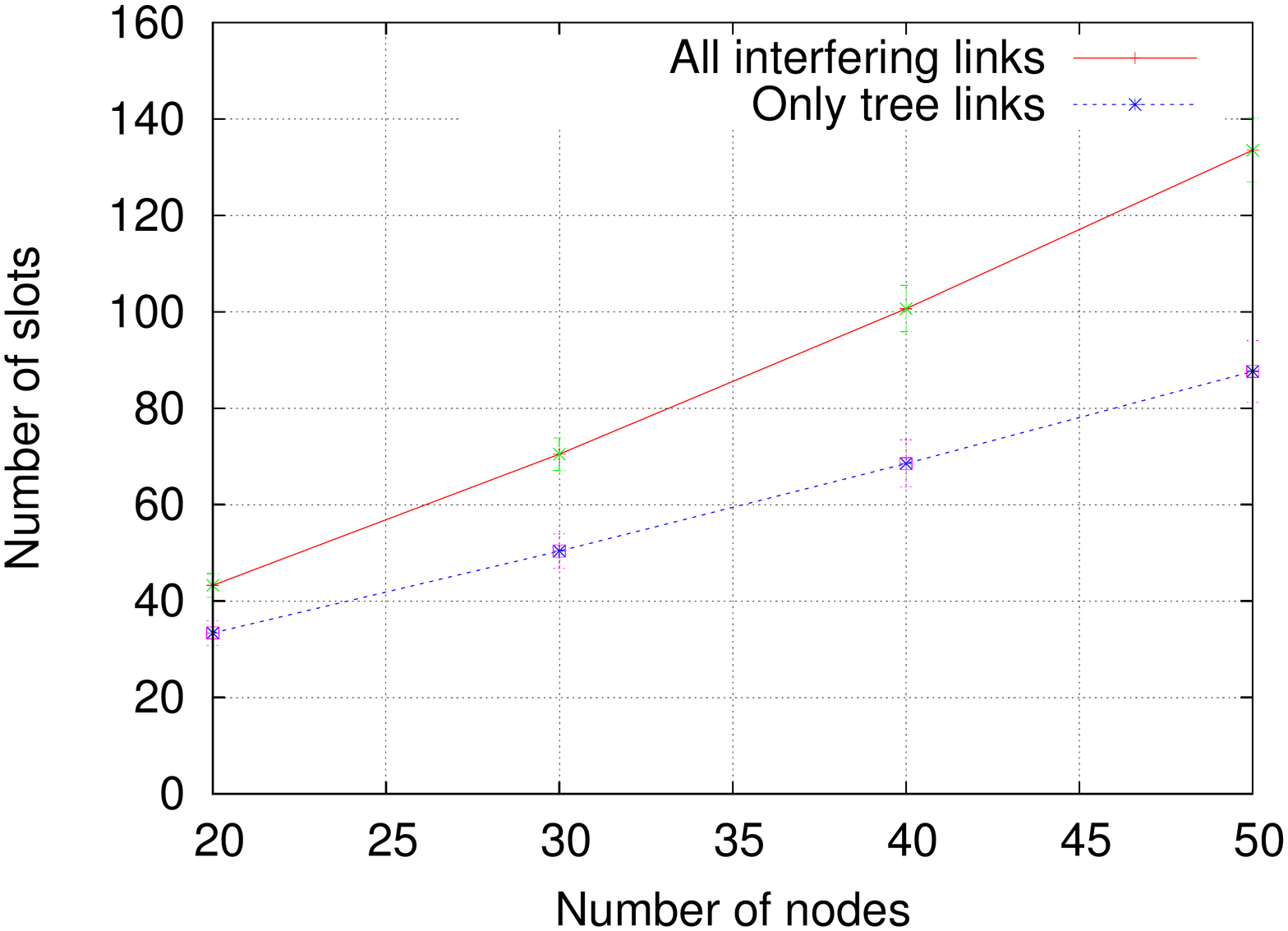}\label{nbSlotV}}
\subfigure[]{\includegraphics[width=1.662in, angle=270]{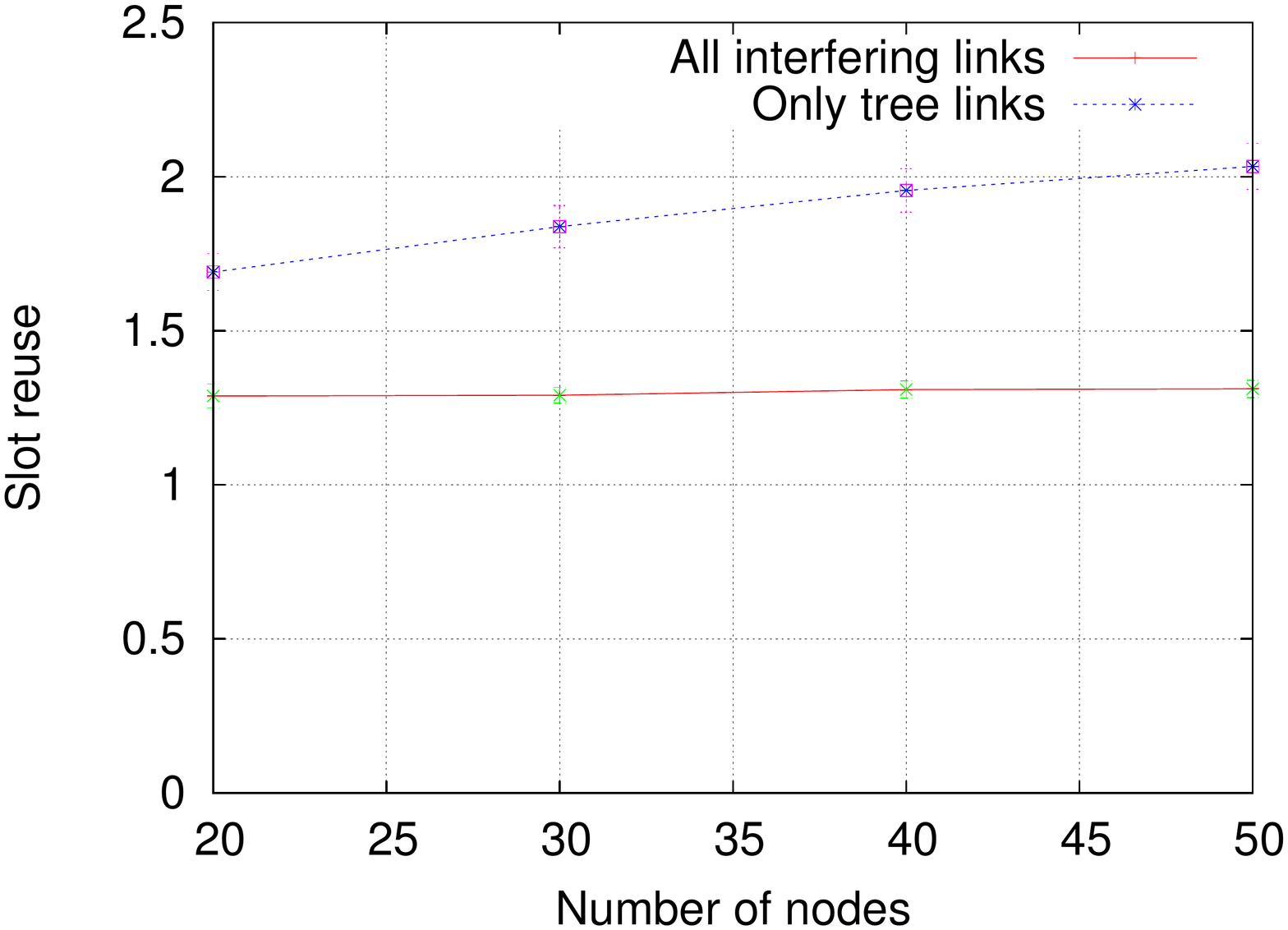}\label{slotReuseV}}
	\caption{Impact of the interference links on: (a) Number of slots; (b) Slot reuse.} \label{impactInterference}
\end{figure}

To evaluate the impact of the interference links on {\em TRASA} performance, we simulate two cases: (1) when all interfering links are considered; (2) when only tree links are considered. As expected, the schedule length in the first case is higher. For instance the number of slots raises from $88$ to $135$ for $50$ nodes. The difference increases with the number of nodes as illustrated in Figure~\ref{nbSlotV}. 
For the same density, any node has more interfering links, so less nodes can be scheduled in the same slot. This justifies the fact that the slot reuse is higher in the second case, as illustrated in Figure~\ref{slotReuseV}.
So, to avoid interferences and achieve collision-free schedule in any real wireless environment, the cost is the schedule length. 

\subsubsection{Impact of the number of nodes, the node data rate and the maximum number of children}
\begin{figure}[!h]
	\centering
\subfigure[]{\includegraphics[width=1.662in, angle=270]{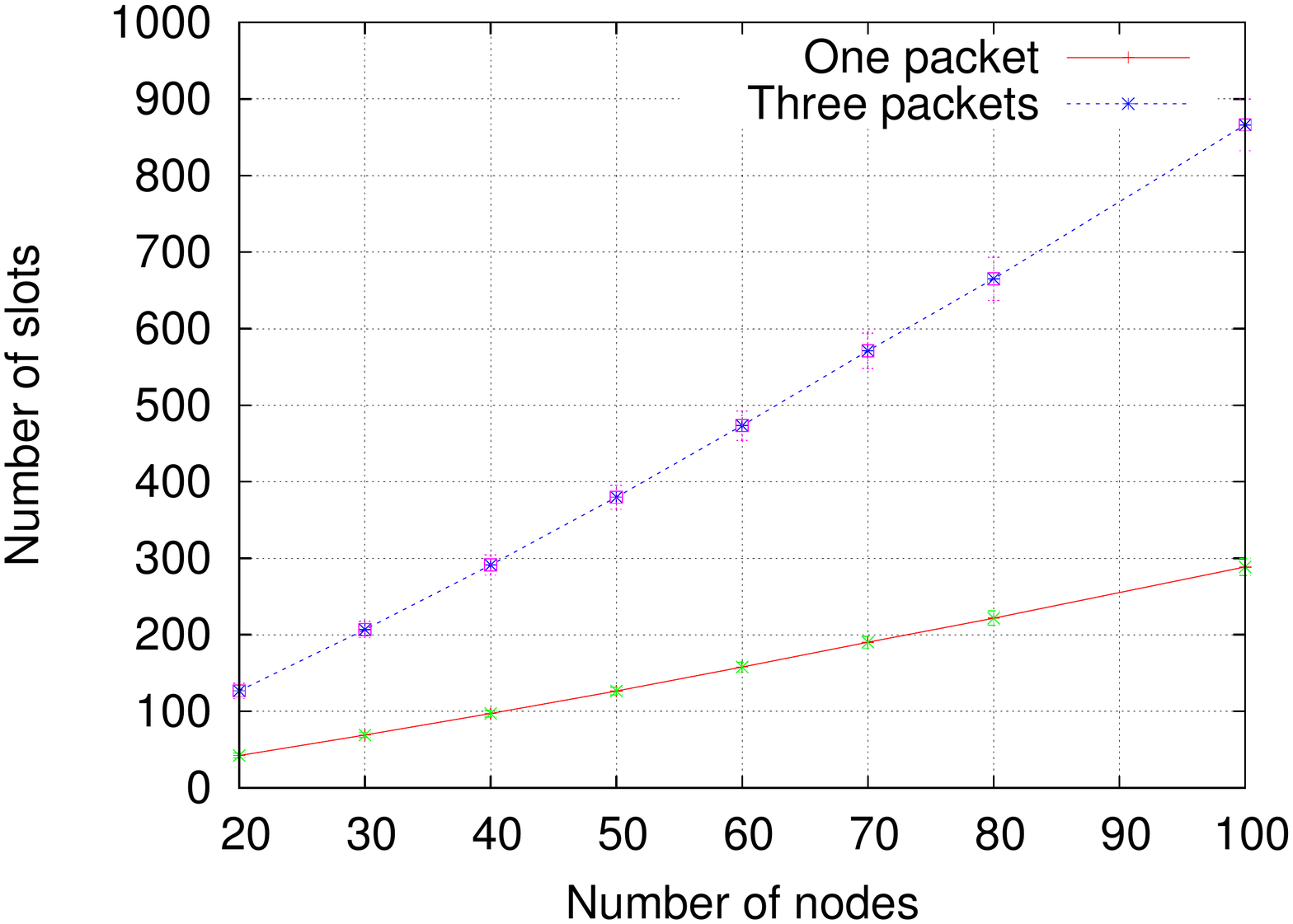}\label{nbSlotRate}}	
\subfigure[]{\includegraphics[width=1.662in, angle=270]{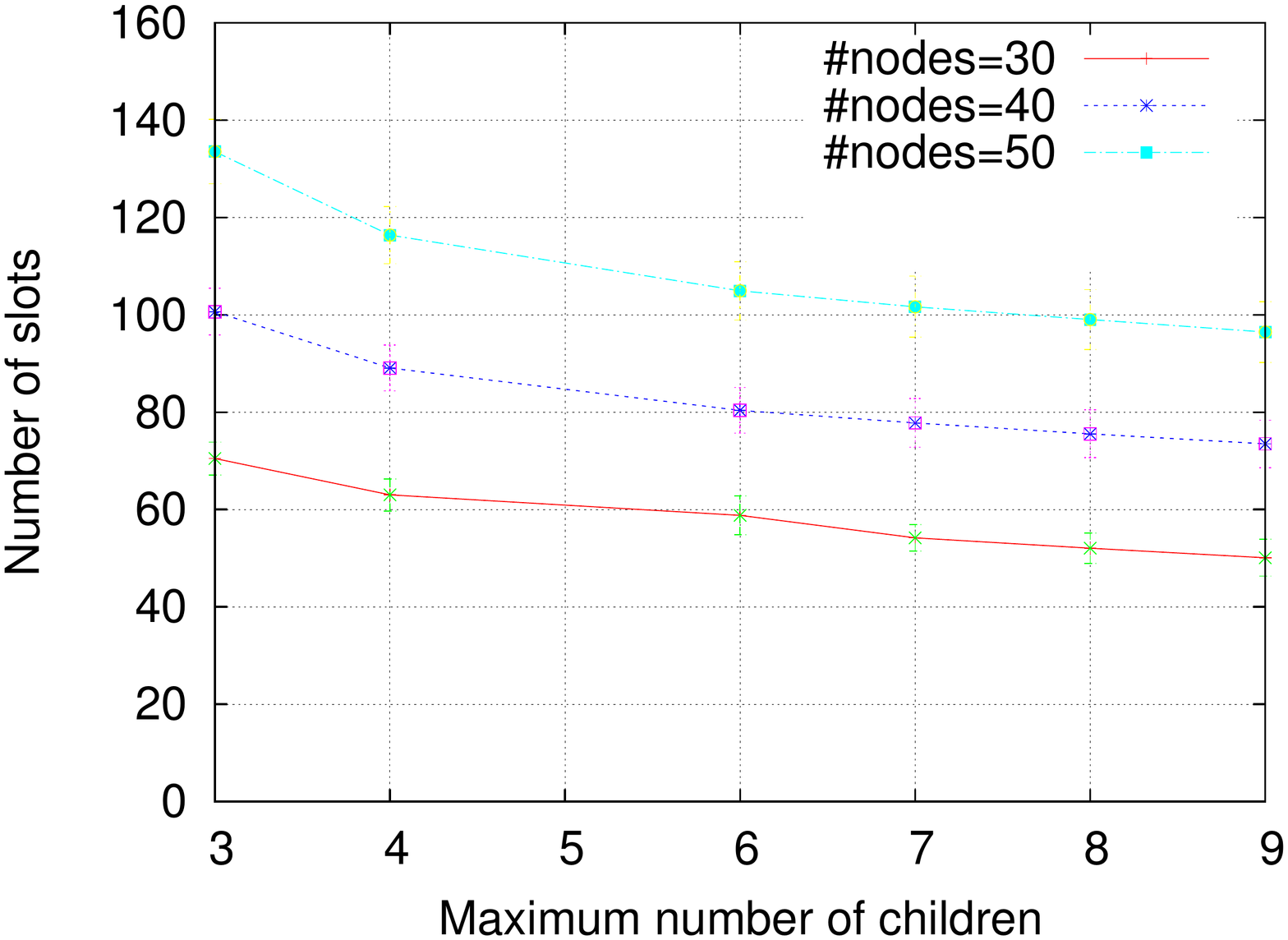}\label{nbSlotForDifferentMAxChildren}}
	\caption{Impact of (a) Data rate (b) Maximum number of children on the number of slots.\label{impactNodeNbRate}}
\end{figure}
For the results illustrated in Figure~\ref{impactNodeNbRate}, the density of nodes (average number of neighbors per node) varies between $4$ and $20$. The Figure~\ref{nbSlotRate} shows that the number of slots increases with the number of nodes. This is because the density increases with the number of nodes, which reduces the spatial reuse of the slots.
However, as the maximum number of children per node increases, the number of slots decreases (see Figure~\ref{nbSlotForDifferentMAxChildren}). Indeed, when the maximum number of children increases, the tree depth decreases. Consequently, the number of slots required by each node which is proportional to its subtree size decreases.
\textit{TRASA} is able to schedule nodes with heterogeneous data rates. Figure~\ref{nbSlotRate} shows that the schedule length increases linearly with the data rate.

\section{Conclusion}\label{conclusion:sec}
In this paper we proved that the time slot assignment problem is NP-complete for any integer $h>1$. We presented \textit{TRASA} an algorithm that takes into account the application requirements to schedule nodes using an optimized number of slots.
We have shown its excellent performance on representative scenarios by simulation.
We plan to extend this work by defining a distributed version of {\em TRASA}.

\end{document}